\documentclass[letter,twocolumn]{jpsj2} %% for letters
\usepackage{bm}
 
\title{{Novel critical exponent of magnetization curves near the ferromagnetic quantum phase transitions of Sr$_{1-x}$A$_{x}$RuO$_{3}$ (A = Ca, La$_{0.5}$Na$_{0.5}$, and La)}
}

\author{\textsc
{Yutaka ITOH}\thanks{E-mail: itoh@kuchem.kyoto-u.ac.jp}
, \textsc{Takao MIZOGUCHI} and 
\textsc{Kazuyoshi YOSHIMURA}
}

\inst{Department of Chemistry, Graduate School of Science, Kyoto University, Kyoto 606-8502}

\abst{
We report a novel critical exponent $\delta \approx$ 3/2 of magnetization curves $M$ = $H^{1/\delta}$ near the ferromagnetic quantum phase transitions of Sr$_{1-x}$A$_{x}$RuO$_{3}$ (A = Ca, La$_{0.5}$Na$_{0.5}$, and La), which the mean field theory of the Ginzburg-Landau-Wilson type fails to reproduce.   
The effect of dirty ferromagnetic spin fluctuations might be a key. 
}

\kword{SrRuO$_3$, ferromagnetic spin fluctuation, quantum critical point, NMR}

\begin{document}
\maketitle
 
Itinerant ferromagnetism has been long studied in physics.
Modern spin fluctuation theory~\cite{Moriya} and 
the concept of quantum critical point~\cite{Hertz,Sachdev} have developed our understanding the metallic magnetism.   
However, further ingredients  e.g. local criticality, quantum protectorate, and hidden criticality
are now required to understand actual experimental results within the present theoretical framework~\cite{Schofield}.  
The itinerant ferromagnetism has still provided us challenging problems. 

SrRuO$_3$ is a perovskite structure with slightly orthorhombic distortion and an itinerant ferromagnet with the Curie temperature $T_\mathrm{C}$ = 160 K~\cite{Goodenough}. 
Ca substitution for Sr ions decreases $T_\mathrm{C}$ and changes the ferromagnetic state into a paramagnetic one at $x\sim$ 0.7 in Sr$_{1-x}$Ca$_x$RuO$_3$~\cite{Kan,Tsuda,Yoshimura,KiyamaMH}. 
Although a ferromagnetic-to-antiferromagnetic transition had been thought to occur at $x \sim$ 0.7 because of the sign change in the Weiss temperature of Curie-Weiss spin susceptibility, 
it turned out that CaRuO$_3$ is a nearly ferromagnetic metal 
and then the $x \sim$ 0.7 transition is a ferromagnetic quantum phase transition~\cite{Yoshimura}. 
The Ru$^{4+}$ (4$d^{4}$) $t_{2g}$-band filling should be invariant with the Ca$^{2+}$ substitution for Sr$^{2+}$, so that the chemical pressure may change the band width and then the electron correlation.

La$_{0.5}$Na$_{0.5}$ and single La ions are known to substitute for Sr ions and to suppress the ferromagnetism~\cite{Kan,HeCava,BW,Nakatsugawa}. 
For A$^{2+}$ = La$^{3+}_{0.5}$Na$^{+}_{0.5}$  and La$^{3+}$ substitution for Sr$^{2+}$ 
in Sr$_{1-x}$A$_{x}$RuO$_{3}$ (A = Ca, La$_{0.5}$Na$_{0.5}$, and La),
one may expect charge disordering~\cite{HeCava} and electron doping effects on the Ru conduction band, respectively. 
The critical concentration $x_\mathrm{c}$ leading to $T_\mathrm{C}$ = 0 K is 0.35 for the La substitution~\cite{BW,Nakatsugawa}. 
The suppression of $T_\mathrm{C}$ is the most steepest for La, secondary for  La$_{0.5}$Na$_{0.5}$ and the slowest for Ca substitution.  
The electron carrier doping is the most effective to suppress the ferromagnetism.  

In this Letter, we report unusual magnetic field dependence of magnetization near the ferromagnetic quantum phase transitions of Sr$_{1-x}$A$_{x}$RuO$_{3}$ (A = Ca, La$_{0.5}$Na$_{0.5}$, and La).
We found a novel critical exponent $\delta \approx$ 3/2 of magnetization curves $M$ = $H^{1/\delta}$, 
in contrast to $\delta$ = 3 of the mean field theory. 
The linear relation of Arrott plot breaks down near all the quantum phase transitions for three types of substitution. 
The dynamic scaling tested by $^{23}$Na NMR suggests the effect of the short mean free path of conduction electrons.    
\begin{figure}[bp]
	\begin{center}
		\includegraphics[width=6.7cm, clip]{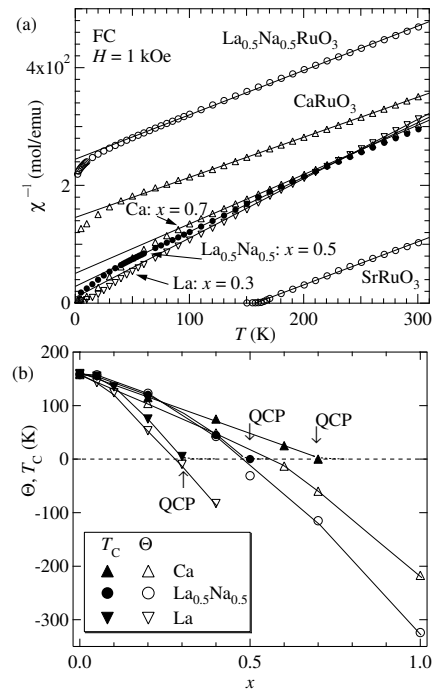}
	\end{center}
	\caption{(a)Inverse magnetic susceptibility $\chi^{-1}$ of Sr$_{1-x}$A$_{x}$RuO$_{3}$ (A = Ca, La$_{0.5}$Na$_{0.5}$, and La). Upward triangles, circles and downward triangles are the data for A = Ca, La$_{0.5}$Na$_{0.5}$, and La, respectively. 
	The magnetic susceptibility $\chi$ is defined by a magnetization $M$ divided by an applied field $H$ = 1 kOe, $M/H$. 
	The solid lines are the fits by an inverse Curie-Weiss law. 
	(b)Magnetic phase diagram of Sr$_{1-x}$A$_{x}$RuO$_{3}$. Concentration $x$ dependence of Weiss temperature $\Theta$ and Curie temperature $T_\mathrm{C}$.
	}
		\label{fig_XT}
\end{figure}% 

Powder samples of Sr$_{1-x}$A$_x$RuO$_3$ were synthesized by a solid state reaction method.
Sr$_{1-x}$Ca$_x$RuO$_3$ and Sr$_{1-x}$La$_x$RuO$_3$: 
Nominal compositions of SrCO$_3$ (99.99 $\%$), Ru metal (99.99 $\%$), CaCO$_3$ (99.99 $\%$) or preheated La$_2$O$_3$ were mixed, ground, pelletized and fired at 900 $^\circ$C in 24 hrs. The products were again ground, pelletized and fired at 900 $^\circ$C in 24 hrs, 
and finally fired at 1300 $^\circ$C in 24 hrs and then cooled down to room temperature in a furnace.   
Sr$_{1-x}$(La$_{0.5}$Na$_{0.5}$)$_x$RuO$_3$: 
For 0$ < x \leq$ 0.5, nominal SrCO$_3$ (99.99 $\%$), RuO$_2$ (99.99 $\%$), NaCO$_3$ (99.99 $\%$) and preheated La$_2$O$_3$ were mixed, ground, pelletized and fired at 750 $^\circ$C in 24 hrs. The products were again ground, pelletized and fired at 900 $^\circ$C in 24 hrs, and finally fired at 1300 $^\circ$C in 24 hrs.  
For $x$ = 0.7 and 1.0, the final heat treatment was done at 1100 $^\circ$C in 24 hrs~\cite{HeCava}. 
Powder X-ray diffraction patterns indicated that all the samples are in a single phase.
   
Magnetization up to 5 T were measured by a SQUID magnetometer (Quantum Design Co. MPMS). 
Figure \ref{fig_XT} (a) shows inverse magnetic susceptibility $\chi^{-1}$ of Sr$_{1-x}$A$_{x}$RuO$_{3}$ (A = Ca, La$_{0.5}$Na$_{0.5}$, and La) at $H$ = 1 kOe. 
The magnetic susceptibility $\chi$ is defined by a magnetization $M$ divided by an applied field $H$ = 1 kOe, $M/H$. 
Solid lines are the fits by an inverse Curie-Weiss law. The Curie constants are nearly the same as that of pure SrRuO$_3$. 
The Weiss temperature changes from positive to negative values with the A ion substitution. 

Figure \ref{fig_XT} (b) shows magnetic phase diagram of Sr$_{1-x}$A$_{x}$RuO$_{3}$. 
The Weiss temperature $\Theta$ and Curie temperature $T_\mathrm{C}$ are plotted as a function of concentration $x$. 
The critical concentrations $x_c$ of $T_\mathrm{C}$ = 0 are $\approx$ 0.7, 0.5 and 0.3 for A = Ca, La$_{0.5}$Na$_{0.5}$, and La, respectively.

Here, we briefly explain the typical magnetization process of a magnetic metal.~\cite{Arrott} 
Classical free energy of the Ginzburg-Landau-Wilson type is given by,
\begin{eqnarray}
F(M)=-g\mu_\mathrm{B}HM+\frac{1}{2\chi_{s}}M^{2}+\frac{1}{4}F_{1}M^{4}+\ldots
\label{eq:free} 
\end{eqnarray}
where $\chi_{s}$ and $F_\mathrm{1}$ are a spin susceptibility and a mode-mode coupling constant.   
Up to 4th order expansion, $\partial F/\partial M$ = 0 gives a thermal equilibrium state.
That is
\begin{equation}
M^{2} = a + b \frac{H}{M},
\label{eq:Arr}
\end{equation}
where $a$=$-1/\chi_{s}F_\mathrm{1}$ and $b$=$g\mu_\mathrm{B}$/$F_\mathrm{1}$.  
The plot of $M^{2}$ against $H/M$ is shortly called Arrott plot.   
Since $a$ = 0 at $T$ = $T_\mathrm{C}$, we obtain
\begin{equation}
M = b^{-1/3}H^{1/3}. 
\label{eq:QCP}
\end{equation}
In general, just at a critical point, we define
\begin{equation}
M\propto H^{1/\delta}. 
\label{eq:CE}
\end{equation}
The mean field theory gives the critical exponent $\delta$ = 3. 

\begin{figure}[tbp]
	\begin{center}
		\includegraphics[width=8.5cm, clip]{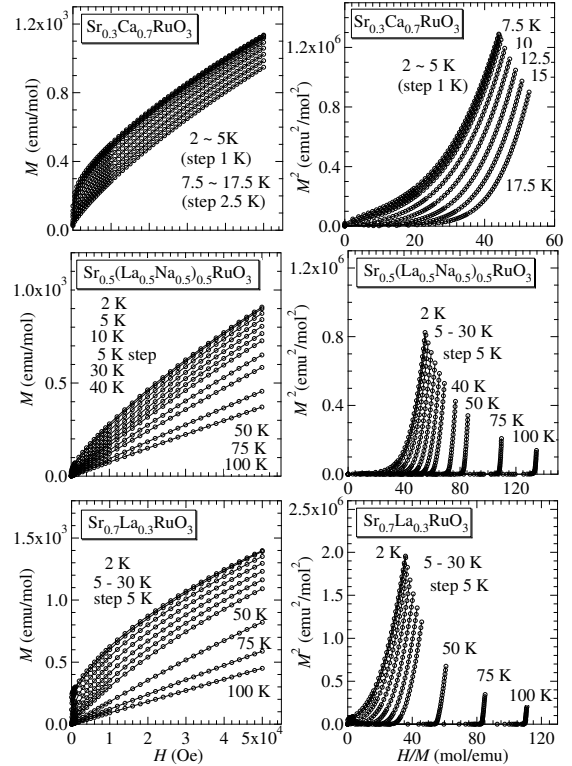}
	\end{center}
	\caption{Field dependence of magnetization $M$ vs $H$ (left panels) and Arrott plots $M^2$ vs $H/M$ (right panels) near the ferromagnetic quantum phase transitions
	of Sr$_{1-x}$A$_{x}$RuO$_{3}$ with A = Ca ($x$ = 0.7), La$_{0.5}$Na$_{0.5}$ ($x$ = 0.5), and La ($x$ = 0.3) from the top to the bottom panels. The Arrott plots show concave curves.  
	}
		\label{fig_MHA}
\end{figure}%
The left panels in Fig. \ref{fig_MHA}  show magnetization curves (field dependence of magnetization) and the right panels show Arrott plots $M^2$ against $H/M$ near the ferromagnetic quantum phase transitions of Sr$_{1-x}$A$_{x}$RuO$_{3}$ with A = Ca ($x$ = 0.7), La$_{0.5}$Na$_{0.5}$ ($x$ = 0.5), and La ($x$ = 0.3) from the top to the bottom panels.	
Obviously, the linearity in the Arrott plots breaks down. 
The Arrott plots show concave curves.
We have observed these behaviors not only at $x_c$ but also around $x_c$
in the individual substitution systems.   
As to Sr$_{1-x}$Ca$_{x}$RuO$_{3}$, the concave curved Arrott plots have already been seen up to high magnetic field of $H$ = 44 T.~\cite{KiyamaMH}

\begin{figure}[tbp]
	\begin{center}
		\includegraphics[width=8.5cm, clip]{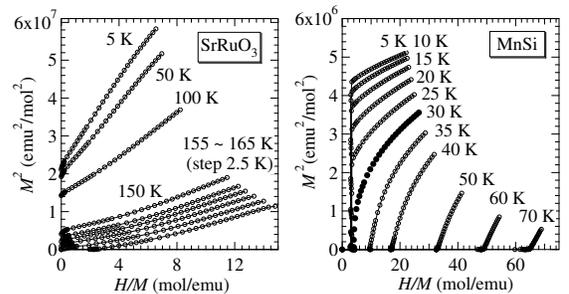}
	\end{center}
	\caption{Arrott plots $M^2$ vs $H/M$ of SrRuO$_3$ (left panel) and MnSi (right panel). The magnetization curves of MnSi are shown as the known example of the breakdown of the linearity in Arrott plots.  The Arrott polts of MnSi show convex curves. 
	}
		\label{fig_MnSi}
\end{figure}%
Figure \ref{fig_MnSi} shows Arrott plots $M^2$ against $H/M$ of SrRuO$_3$ and MnSi.~\cite{Yasuoka}
The Arrott plots hold relatively well for SrRuO$_3$.
That is, the critical exponent $\delta$ near $T_\mathrm{C}\approx$ 160 K is close to the mean field value 3~\cite{single}. 
The Arrott polts of MnSi show convex curves~\cite{Jaccarino}. 
This is an another breakdown of Arrott plots different from Sr$_{1-x}$A$_x$RuO$_3$ with $x_c$.  
The linear relation is seen in the plot of $M^4$ against $H/M$~\cite{TakahashiMH1}.

We estimated ``$b$"$ \equiv \Delta M^2/\Delta (H/M)$ as a function of  magnetic field $H$ for each Arrott plot.  
For each substitution system, we observed a crossover from a nearly $H$-independent ``$b$"
for $x$ = 0 at $T_\mathrm{C}$ to ``$b$"$ \propto H$ for $x$ = $x_c$ at low temperatures ($T\geq$ 2 K $> T_\mathrm{C}$). 
Thus, we found an empirical relation ``$b$"$ \propto H$ near the quantum phase transitions, 
although $a$ and $b$ must not depend on $H$ in eq. (\ref{eq:Arr}).  
This is equivalent to $M\propto H^{2/3}$.
The equation of $M = pH^{2/3}+q$ with the $T$- and $x$-dependent $p$ and $q$ was applied to the magnetization around $x$ = $x_c$. The $q$ vanished in $x$ = $x_c$ at low temperatures.

\begin{figure}[tbp]
	\begin{center}
		\includegraphics[width=6.7cm, clip]{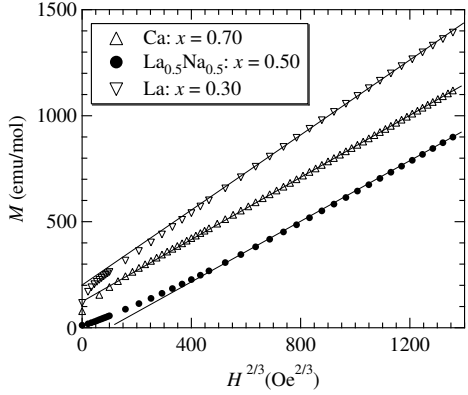}
	\end{center}
	\caption{Magnetization $M$ plotted against $H^{2/3}$ at $T$ = 5 K near the ferromagnetic quantum phase transitions of Sr$_{1-x}$A$_{x}$RuO$_{3}$ with A = Ca ($x$ = 0.7), La$_{0.5}$Na$_{0.5}$ ($x$ = 0.5), and La ($x$ = 0.3).
	The straight lines are visual guides. 	
	}
		\label{fig_MHS}
\end{figure}

In Fig. \ref{fig_MHS}, the magnetizations $M$ are plotted against $H^{2/3}$ at $T$ = 5 K 
near the ferromagnetic quantum phase transitions of Sr$_{1-x}$A$_{x}$RuO$_{3}$ 
with A = Ca ($x$ = 0.7), La$_{0.5}$Na$_{0.5}$ ($x$ = 0.5), and La ($x$ = 0.3).
The linear relations hold well. 
The critical exponent of magnetization is $\delta\approx$ 3/2 for $x_c$ at lower temperatures.  
The finite temperature scaling relation could not apply to the $x$ = $x_c$ sample because of $T_\mathrm{C}$ = 0 K. 
The test for a scaling relation with respect to $g$ = $(x - x_c)/x_c$ extrapolated toward $T$ = 0 K ~\cite{Gauss} remains to be a problem.  

MnSi and BaRuO$_3$~\cite{BaRuO3} show the critical exponent $\delta\approx$ 5,
which can be understood by the effects of thermal ferromagnetic spin fluctuations on the magnetization curves~\cite{TakahashiMH1,TakahashiMH2}. 
The Gaussian fluctuation effects on the magnetization leads to a small $\delta$ = 7/3 $\approx$ 2.33~\cite{Gauss}, which may apply to BaIrO$_3$~\cite{BaIrO3}.
The $\delta \approx$ 1.5 of Sr$_{1-x}$A$_{x}$RuO$_{3}$ smaller than the mean field value $\delta$ = 3 could not be reproduced by the higher order terms of magnetization nor by the Gaussian fluctuations.
First order like inhomogeneous quantum phase transition~\cite{Daniel,Uemura} might be associated with the small $\delta$.
Some effects on magnetization process via spin fluctuations might affect $\delta$.
 
In order to study the microscopic spin fluctuation spectrum, we performed $^{23}$Na (nuclear spin $I$ = 3/2 and nuclear gyromagnetic ratio $\gamma_{\rm n}/2\pi$ = 11.262 MHz/T) NMR spin-echo measurements
for Sr$_{1-x}$(La$_{0.5}$Na$_{0.5}$)$_{x}$RuO$_3$ with $x$ = 0.5 and 1.0 at $H$ = 7.48414 T. 
\begin{figure}[htbp]
	\begin{center}
		\includegraphics[width=6.7cm, clip]{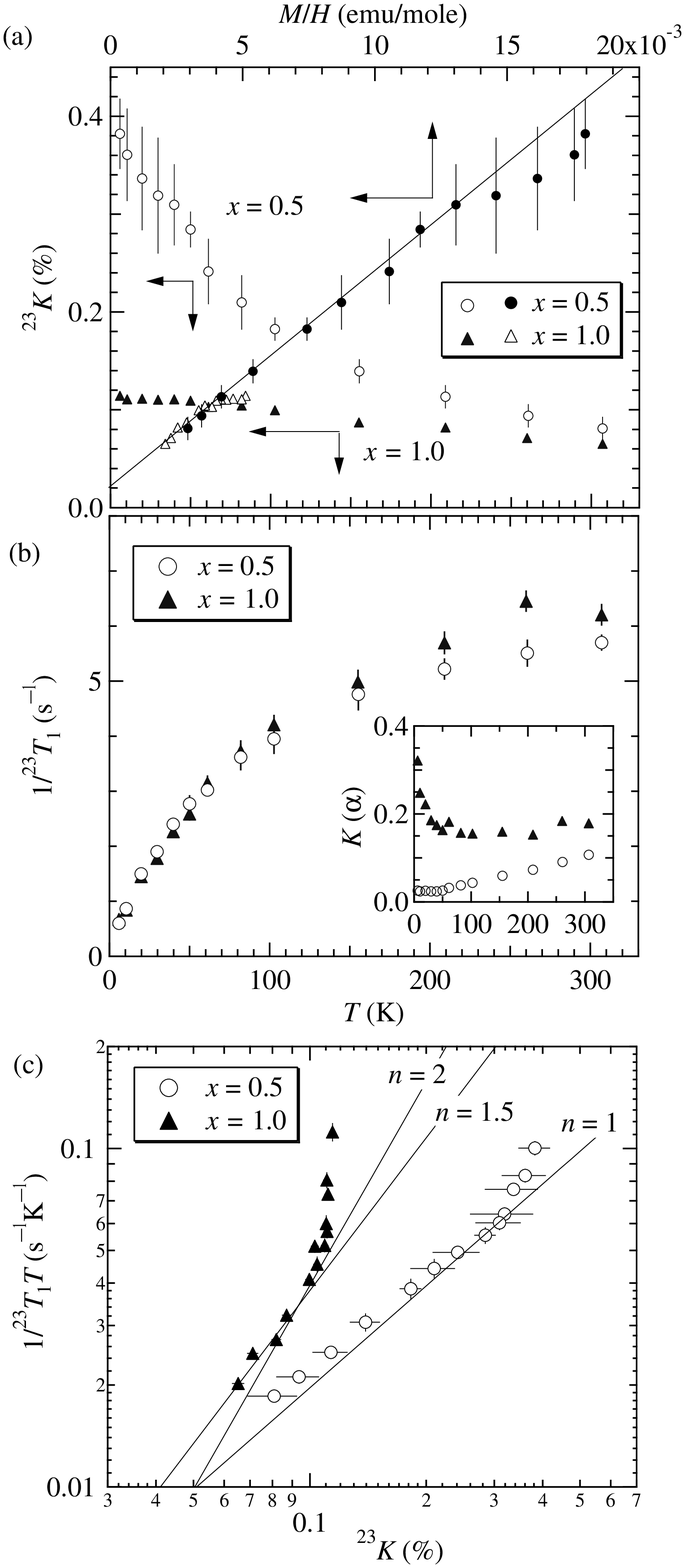}
	\end{center}
	\caption{(a)$^{23}$Na Knight shifts plotted against temperature (bottom axis) and against magnetic susceptibility defined by $M$/$H$ at $H$ = 5 T (top axis) for Sr$_{1-x}$(La$_{0.5}$Na$_{0.5}$)$_{x}$RuO$_3$ with $x$ = 0.5 (circles) and 1.0 (triangles). The solid line is a linear function fit to the $K-\chi$ plot. 
	(b)Temperature dependence of $^{23}$Na nuclear spin-lattice relaxation rate 1/$T_1$ for Sr$_{1-x}$(La$_{0.5}$Na$_{0.5}$)$_{x}$RuO$_3$ with $x$ = 0.5 and 1.0. The inset shows temperature dependence of $K(\alpha)$. 
	(c)Log-log plots of $^{23}$Na nuclear spin-lattice relaxation rate divided by temperature 1/$T_{1}T$ against Knight shift $K$ for Sr$_{1-x}$(La$_{0.5}$Na$_{0.5}$)$_{x}$RuO$_3$ with $x$ = 0.5 and 1.0. The solid lines are visual guides for $K^{n}$ with $n$ = 1, 1.5 and 2
	}
		\label{fig_NMR}
\end{figure}%

$^{23}$Na NMR frequency spectra systematically changed with La$_{0.5}$Na$_{0.5}$ substitution. 
Figure \ref{fig_NMR} (a) shows $^{23}$Na Knight shifts against temperature (bottom axis) and against magnetic susceptibility defined by $M$/$H$ at $H$ = 5 T (top axis) for $x$ = 0.5 and 1.0.
The temperature dependent spin part of Knight shift $K$ is proportional to the spin susceptibility $\chi_s$ via a hyperfine coupling constant $A_\mathrm{hf}$, $K_s(T) = A_{\rm hf}\chi_s(T)/N_{\rm A}\mu_{\rm B}$ ($N_{\rm A}$ is the Avogadro number and $\mu_{\rm B}$ is the Bohr magneton).
From the $K-\chi$ plots and the linear function fit, $A_\mathrm{hf}$ was estimated to be +1.2$\pm$ 0.1 kOe/$\mu_\mathrm{B}$.  

Figure \ref{fig_NMR} (b) shows temperature dependence of $^{23}$Na nuclear spin-lattice relaxation rate 1/$T_1$ for $x$ = 0.5 and 1.0 measured by an inversion recovery spin-echo technique. 
For both $x$ = 0.5 and 1.0,
1/$T_1$ shows non-Korringa behavior and levels off at high temperatures.
The critical slowing down of 1/$T_1 \propto T^{-1/3}$ just at the quantum critical point~\cite{IshigakiMoriya} is not observed for $x$ = 0.5.
This is consistent with the absence of the critical divergence for the Ca substituted $x \sim$ 0.7~\cite{Yoshimura,Uemura}.
No divergence in Fig. \ref{fig_NMR} (b), however, might be due to the magnetic field $\sim$ 7.5 T.

The inset in Fig. \ref{fig_NMR} (b) shows temperature dependence of $K(\alpha)$ = $S/T_1TK_s^2$ 
($S \equiv \gamma_e\hbar/4\pi k_{\rm B}\gamma_{\rm n}$,
the electron gyromagnetic ratio $\gamma_e$ and an exchange enhancement factor $\alpha$)~\cite{MoriyaT1}.
Exchange-enhancement effects deviate the Korringa ratio from that of free electron gas by $K(\alpha)$~\cite{MoriyaT1}.
The ferromagnetic and antiferromagnetic spin fluctuations give $K(\alpha) <$ 1 and $>$ 1, respectively. 
The $K(\alpha) <<$ 1 in the inset of Fig. \ref{fig_NMR} (b) evidences that both $x$ = 0.5 and 1.0 systems are nearly ferromagnetic metals.   

For a nearly ferromagnetic metal, the dynamical spin susceptibility $\chi\prime\prime(q, \omega)$ with dynamic critical exponent $z$
and magnetic correaltion length $\xi$ is given by a scaling hypothesis,
\begin{equation}
\chi\prime\prime(q, \omega) = \xi^{2}f(\xi q, \xi^{z}\omega),
\label{eq:scalingXqw}
\end{equation}
where $f$ is an appropriate function. 
One should note $z$ = 3 and 4 in the clean and dirty limits, respectively~\cite{Hertz,Millis}. 
This is because in the dirty limit, the quasi-particle life time  and the mean free path are short and then the spin fluctuation spectrum is diffusive,
that is the characteristic frequency $\Gamma (q)\propto q^2$ ($q^4$ at $T$ = $T_{\rm C}$), 
being in contrast to the clean $\Gamma (q)\propto q$ ($q^3$ at $T$ = $T_{\rm C}$)~\cite{Fulude}.  
For the nearly ferromagnetic metal in $D$(= 2 and 3) dimensions, the nuclear spin-lattice relaxation rate divided by temperature 1/$T_1T$ is expressed by the static spin susceptibility $\chi_s\propto \xi^2$ and the Knight shift $K_s\propto \chi_s$,
\begin{equation}
\frac{1}{T_1T} \propto \xi^{2+z-D} \propto K_s^{1+(z-D)/2}.
\label{eq:scalingT1}
\end{equation}
This is a consequence from the dynamic scaling law. 
For a 3$D$ nearly ferromagnetic metal, we obtain 1/$T_1T\propto K$ in the clean limit~\cite{MU}. 
In the dirty limit, we obtain 1/$T_1T\propto K^{1.5}$ for $z$ = 4. 

In a Lorentzian model, the $\chi\prime\prime(q, \omega)$ is characterized by 
the energy width $T_{\rm 0}$ of spin fluctuations and the spread $T_{\rm A}$ in the $q$-space~\cite{TakahashiSCR}.
The nuclear spin-lattice relaxation rate 1/$T_1$ for the 3$D$ nearly ferromagnetic metal in the clean limit is given by
\begin{equation}
\frac{1}{T_1} = \frac{3\hbar\gamma_{\rm n}^2A_{\rm hf}^2}{4\pi}\frac{t}{T_{\rm A}y},
\label{eq:SCRT1}
\end{equation}
where the reduced temperature $t$ = $T$/$T_{\rm 0}$ and the reduced magnetic susceptibility 1/$y$ = $(\xi q_{\rm B})^2$ ($q_{\rm B}$ is the effective spherical radius of the Brillouin zone)~\cite{IshigakiMoriya}. 
The static spin susceptibility is expressed by $\chi_s/N_{\rm A}$ = 1/$2T_{\rm A}y$.
The self-consistent renormalization (SCR) theory reproduces the Curie-Weiss behavior of 1/$y$
as a function of $t$ and the distance $y_{\rm 0}$ to the quantum critical point~\cite{IshigakiMoriya}. 
At $T\rightarrow$ 0, 1/$T_1\propto t/T_{\rm A}y_{\rm 0}$ is Korringa like.
At high temperatures $t >$ $y_{\rm 0}T_{\rm 0}$, $t/y$ levels off and takes $\sim$ 27~\cite{IshigakiMoriya,MU}. 
Thus, we obtain
\begin{equation}
\frac{1}{T_1} \rightarrow \frac{81\hbar\gamma_{\rm n}^2A_{\rm hf}^2}{4\pi}\frac{1}{T_{\rm A}}.
\label{eq:HighT1}
\end{equation}
The temperature dependences of 1/$T_1$ in Fig. \ref{fig_NMR} (b) for A = La$_{0.5}$Na$_{0.5}$ and the previous ones for A = Ca~\cite{Yoshimura} are consistent with the SCR results except the critical slowing down toward $T$ = 0 K at the quantum critical point.

Figure \ref{fig_NMR} (c) shows log-log plots of $^{23}$Na NMR 1/$T_{1}T$ against Knight shift $K$ with temperature as an implicit parameter for $x$ = 0.5 and 1.0.  
 The relation between 1/$T_{1}T$ and $K$ is not a simple $1/T_1T\propto K$. 
Both $x$ = 0.5 and 1.0 systems may be located in the intermediate region between good and bad metals,
which might be associated with the small exponent $\delta\approx$ 3/2 of magnetization curves.

In conclusion, we found a novel small critical exponent $\delta \approx$ 3/2 of magnetization curves $M\propto H^{1/\delta}$ near the ferromagnetic quantum phase transitions of Sr$_{1-x}$A$_{x}$RuO$_{3}$ (A = Ca, La$_{0.5}$Na$_{0.5}$, and La). 
The $^{23}$Na NMR test for the dynamic scaling law indicated that
the dirty ferromagnetic spin fluctuation spectrum might be a key. 

We thank H. Ohta, T. Waki, W. Zhang and C. Michioka for their experimental supports and helpful discussions,
and S. Miyasaka, M. Hagiwara and I. Terasaki for valuable discussions.  
This work was supported in part by a Grant-in-Aid for Science Research on
Priority Area, ``Invention of Anomalous Quantum Materials," from the Ministry
of Education, Culture, Sports, Science and Technology of Japan (16076210)
and in part by a Grant-in-Aid for Scientific Research from the Japan Society
for Promotion of Science (Grant No. 19350030).


\begin{thebibliography}{99} %% The number "99" means that this list has more than nine items.

\bibitem{Moriya}
T. Moriya: {\it Spin Fluctuations in Itinerant Electron Magnetism} (Springer-Verlag, Berlin 1985).

\bibitem{Hertz}
J. A. Hertz: Phys. Rev. B {\bf 14} (1976) 1165. 

\bibitem{Sachdev}
S. Sachdev: {\it Quantum Phase Transitions} (Cambridge University Press, New York 1999).

\bibitem{Schofield}
A. Schofield: Phys. World Aug. (2003) 23. 

\bibitem{Goodenough}
J. M. Longo, P. M. Raccah and J. B. Goodenough: J. App. Phys. {\bf 39} (1968) 1327.

\bibitem{Kan}
A. Kanbayashi: J. Phys. Soc. Jpn. {\bf 44} (1978) 108. 

\bibitem{Tsuda}
F. Fukunaga and N. Tsuda: J. Phys. Soc. Jpn. {\bf 63} (1994) 3798.

\bibitem{Yoshimura} 
K. Yoshimura, T. Imai, T. Kiyama, K. R. Thurber, A. W. Hunt and K. Kosuge: Phys. Rev. Lett. {\bf 83} (1999) 4397. 

\bibitem{KiyamaMH}
T. Kiyama, K. Yoshimura, K. Kosuge, H. Mitamura and T. Goto: J. Phys. Soc. Jpn. {\bf 68} (1999) 3372.

\bibitem{HeCava}
T. He, Q. Huang and R. J. Cava: Phys. Rev. B {\bf 63} (2000) 024402.

\bibitem{BW}
R. J. Bouchard and J. F. Weiher: J. Solid State Chem. {\bf 4} (1972) 80. 

\bibitem{Nakatsugawa}
H. Nakatsugawa, E. Iguchi and Y. Oohara: J. Phys.: Condens. Matter {\bf 14} (2002) 415. 

\bibitem{Arrott}
A. Arrott: Phys. Rev. {\bf 108} (1957) 1394. 

\bibitem{Yasuoka}
H. Yasuoka, V. Jaccarino, R. C. Sherwood and J. H. Wernick: J. Phys. Soc. Jpn. {\bf 44} (1978) 842.
The $MH$ curves have been measured for the MnSi sample in this report.
We thank Prof. Yasuoka for kind supply of the sample. 

\bibitem{single}
D. Kim, B. L. Zink, F. Hellman, S. McCall, G. Cao and J. E. Crow: Phys. Rev. B {\bf 67} (2003) R100406. 

\bibitem{Jaccarino}
D. Bloch, J. Voiron, V. Jaccarino and J. H. Wernick: Phys. Lett. {\bf 51} (1975) 259.

\bibitem{TakahashiMH1}
Y. Takahashi: J. Phys. Soc. Jpn. {\bf 55} (1986) 3553.

\bibitem{TakahashiMH2}
Y. Takahashi: J. Phys.: Condens. Matter {\bf 13} (2001) 6323. 

\bibitem{Gauss}
M. A. Continentino: {\it Quantum Scaling in Many-Body Systems} (World Scientific, Singapore 2001).

\bibitem{BaRuO3}
J.-S. Zhou, K. Matsubayashi, Y. Uwatoko, C.-Q. Jin, J.-G. Cheng, J. B. Goodenough, Q. Q. Liu, T. Katsura, A. Shatskiy and E. Ito: Phys. Rev. Lett. {\bf 101} (2008) 077206. 

\bibitem{BaIrO3}
T. Kida, A. Senda, S. Yoshii, M. Hagiwara, T. Takeuchi, T. Nakano and I. Terasaki: in press in Euro. Phys. Lett.  

\bibitem{Daniel} 
M. Daniel, J. I. Budnick, W. A. Hines, Y. D. Zhang, W. G. Clark and A. R. Moodenbaugh: J. Phys.: Condens. Matter {\bf 12} (2000) 3857.

\bibitem{Uemura} 
Y. J. Uemura {\it et al.}: Nature Phys. {\bf 31} (2007) 29.  

\bibitem{IshigakiMoriya}
A. Ishigaki and T. Moriya: J. Phys. Soc. Jpn. {\bf 65} (1996) 3402. 

\bibitem{MoriyaT1}
T. Moriya: J. Phys. Soc. Jpn. {\bf 18} (1963) 516.

\bibitem{Millis}
A. J. Millis: Phys. Rev. B {\bf 48} (1993) 7183. 

\bibitem{Fulude}
P. Fulude and A. Luther: Phys. Rev. {\bf 170} (1968) 570.

\bibitem{MU}
T. Moriya and K. Ueda: Solid State Commun. {\bf 15} (1974) 169. 

\bibitem{TakahashiSCR}
Y. Takahashi and T. Moriya: J. Phys. Soc. Jpn. {\bf 54} (1985) 1592.  

\end{thebibliography}
\end{document}